\documentclass[11pt,preprint]{aastex}
\pdfoutput=1
\usepackage{bm}
\usepackage{graphicx}
\usepackage{footnote}
\usepackage{color}
\usepackage{amsmath}
\usepackage{mathtools}
\usepackage{multirow}
\usepackage{pdflscape}

\usepackage{afterpage}

\newcommand\msun {M_\odot}
\newcommand\mearth {{M_\oplus}}
\newcommand{\mathbold}[1]{\mbox{\boldmath $\bf#1$}}

\newcommand\mubold{{\mathbold \mu}}

\newcommand\ltsima{$\; \buildrel <\over\sim \;$}
\newcommand\simlt{\lower.5ex\hbox{\ltsima}}
\newcommand\gtsima{$\; \buildrel >\over\sim \;$}
\newcommand\simgt{\lower.5ex\hbox{\gtsima}}

\shorttitle{}
\shortauthors{Bhattacharya et al}

\begin{document}

\title{The Star Blended with the MOA-2008-BLG-310 Source Is Not the Exoplanet Host Star}


\author{A.~Bhattacharya\altaffilmark{1,2},
D.P.~Bennett\altaffilmark{1,2}, J.~Anderson\altaffilmark{3}, I.A.~Bond\altaffilmark{4}, A.~Gould\altaffilmark{5}, V.~Batista\altaffilmark{6}, J.P.~Beaulieu\altaffilmark{6}, P.~Fouqu\'e\altaffilmark{7}, J.B~Marquette\altaffilmark{6}, R.~Pogge\altaffilmark{5}}

\altaffiltext{1}{Department of Physics,
    University of Notre Dame, 225 Nieuwland Science Hall, Notre Dame, IN 46556, USA; 
    Email: {\tt aparna.bhattacharya@nasa.gov}}
\altaffiltext{2}{Code 667, NASA Goddard Space Flight Center, Greenbelt, MD 20771, USA}
\altaffiltext{3}{Space Telescope Institute, 3700 San Martin Drive, Baltimore, MD 21218, USA}
\altaffiltext{4}{Institute of Natural and Mathematical Sciences, Massey University, Auckland 0745, New Zealand}
\altaffiltext{5}{Department of Astronomy, Ohio State University, 140 W. 18th Ave., Columbus, OH 43210, USA}
\altaffiltext{6}{UPMC-CNRS, UMR 7095, Institut d’Astrophysique de Paris, 98Bis Boulevard Arago, F-75014 Paris}
\altaffiltext{7}{Canada France Hawaii Telescope, 65-1238 Mamalahoa Hwy, Waimea, HI 96743}

\begin{abstract}
High resolution Hubble Space Telescope ({\it HST})
image analysis of the MOA-2008-BLG-310 microlens system indicates that the 
excess flux at the location of the source found in the discovery paper cannot primarily be due to the lens star
because it does not match the lens-source relative proper motion, $\mu_{\rm rel}$, predicted by
the microlens models. This excess flux is most likely to be due to an unrelated star that happens to
be located in close proximity to the source star. Two epochs of  {\it HST} observations indicate proper
motion for this blend star that is typical of a random bulge star, but is not consistent with a 
companion to the source or lens stars if the flux is dominated by only one star, aside from the
lens. We consider models in which the excess flux is due
to a combination of an unrelated star and the lens star, and this yields 95\%
confidence level upper limit on the
lens star brightness of $I_L > 22.44$ and $V_L >23.62$. A Bayesian analysis using a standard
Galactic model and these magnitude limits yields a host star mass $M_h = 0.21 ^{+0.21} _{-0.09}~ \msun$,
a planet mass of $m_p = 23.4 ^{+23.9} _{-9.9}~\mearth$ at a projected separation of 
$a_\perp = 1.12^{+0.16}_{-0.17},$AU.
This result illustrates excess flux in a high resolution image of a microlens-source system need
not be due to the lens. It is important to check that the lens-source relative proper motion is
consistent with the microlensing prediction. The high resolution image analysis techniques 
developed in this paper can be used to verify the WFIRST exoplanet microlensing survey
mass measurements.
\end{abstract}
\section{Introduction}
\label{sec-intro}
Most of the known exoplanets have been discovered by the Doppler radial velocity method
\citep{radial_vel1,radial_vel2} or by the transit method \citep{transit2}, with the largest number
coming from the Kepler exoplanet transit mission \citep{kepler1,kepler2}.
Despite the large number of exoplanets discovered, our knowledge about the distribution of 
exoplanets by these methods is limited by the selection effects of these methods. Most of these 
planets are hot or warm planets at small orbital ($\leqslant$1 AU) separation from their host 
stars, and the planets in wider orbits are generally more massive than Saturn. Microlensing is 
the only method that is sensitive to the low mass planets at orbital separations larger than the
snow line. According to the core accretion theory of the planet formation \citep{lissauer_araa},
the planet formation process is most efficient beyond the snowline \citep{snowline2,snowline3} where
the protoplanetary disk is cold enough for ices to condense. This gives a higher density of solid
material that can coagulate to start the planet formation process. Hence, the microlensing
method allows us to study the demographics of the planetary systems in the favored planetary
birthplace, beyond the snow line.

Gravitational microlensing is the method for detecting the exoplanets with masses as low as an Earth mass
\citep{bennett1} at a distance $\sim 1$-$8\,$kpc from earth. Since the technique does not 
depend on the light from the exoplanet or its host star, it is very effective in detecting planets that orbit
very faint stars, including planets orbiting stars in the Galactic bulge. A number of planets with
host star probably in the bulge have already been discovered, including OGLE-2005-BLG-390 \citep{ogle390}, 
OGLE-2008-BLG-092 \citep{ogle092}, OGLE-2008-BLG-355 \citep{ogle355}, 
MOA-2008-BLG-310Lb \citep{moa310},MOA-2009-BLG-319 \citep{moa319}, 
MOA-2011-BLG-262 \citep{bennett14}, MOA-2011-BLG-293 \citep{moa293} and OGLE-2014-BLG-1760Lb \citep{ogle1760}. 

To date, about $\sim$50 planets have been discovered using microlensing, and 30 of these have
been used to derive the exoplanet mass ratio function \citep{suzuki2016}, which
describes the occurrence rate of planets as a function of their mass ratio, $q$, and separation
in Einstein radius units. To extend these results to find
exoplanetary mass as a function of the host star mass and galactocentric distance, we must 
determine the planet and host star mass and their distance from Earth. For most planetary
microlensing light curve events we obtain the planet-host star mass ratio, separation
in Einstein radius, and the source radius crossing time, $t_*$, which leads to a determination
of the angular Einstein radius, $\theta_E$. This is not enough information to measure the 
host star and planet masses, but we can estimate them with a Bayesian analysis using a
Galactic model, with the assumption that the exoplanet mass function doesn't depend on the
mass or distance of the host star. The masses of the planets and host star can be determined for
events that include a measurement of the microlensing parallax effect 
\citep{gould-1994,gould-1995,gould-1999,gaudi-ogle109} 
or by detecting the lens in the high resolution follow up images
\citep{bennett06,ogle169,batistaogle169}. In some cases, both microlensing parallax and 
high resolution follow-up imaging can provide mass measurements by independent methods
\citep{gaudi-ogle109,bennett-ogle109,bennett16,ogle0026}. High resolution image analyses of planetary
microlensing events have already yielded the mass measurements or upper limits
for OGLE-2003-BLG-235 \citep{bennett06}, OGLE-2006-BLG-071Lb \citep{dong-ogle71},
OGLE-2005-BLG-169Lb \citep{ogle169,batistaogle169}, OGLE-2007-BLG-368 \citep{sumi10},
MOA-2007-BLG-192 \citep{moa192_naco}, MOA-2008-BLG-310 \citep{moa310}, 
MOA-2011-BLG-262Lb \citep{bennett14}, MOA-2011-BLG-293 \citep{moa293}, 
OGLE-2012-BLG-0026 \citep{ogle0026}, OGLE-2012-BLG-0563Lb \citep{fukui15}, 
OGLE-2012-BLG-0950Lb \citep{koshimoto16}, and MOA-2013-BLG-605Lb \citep{sumi16}. 
In this paper we take a second look at the exoplanetary microlensing event MOA-2008-BLG-310
with two epochs of {\it Hubble Space Telescope} (HST) imaging taken in 2012 and 2014. While
the discovery paper presented excess starlight at the position of the source with VLT 
adaptive optics (AO) imaging, our HST images allow us to determine if the star or 
stars responsible for this excess flux have a lens-source relative proper motion that is consistent with
the lens (and planetary host) star.

The MOA (Microlensing Observations in Astrophysics) group identified the microlensing 
event MOA-2008-BLG-310 on July 6, 2008 and 2 days later issued a high magnification alert. 
The peak of this event and its planetary anomaly was meticulously covered by $\mu$FUN 
(Microlensing Follow Up Network) in CTIO $I$, $V$, $H$ and $\mu$FUN $R$ and Bronberg 
unfiltered passbands. MiNDSTEp and PLANET collaboration also took data in the $I$ band. 
This event was also observed using VLT/NACO AO system on July 28, 2008, and these data
showed additional $H$-band flux on top of the source, suggesting a possible detection of the
planetary host star. However, as \citep{moa310} pointed out, it is also possible that the
excess flux could be due to a companion to the source, a companion to the lens, or an 
unrelated star. There was no ambiguity in the interpretation of the excess flux for 
planetary microlensing event OGLE-2005-BLG-169 because the high angular resolution
follow-up observations from {\it HST} and Keck were able to demonstrate the lens-source relative
proper motion and measure the host star flux in four passbands \citep{ogle169,batistaogle169}.
In this paper, we present a similar analysis of planetary microlensing event MOA-2008-BLG-310.
It was observed by HST/WFC3-UVIS in the $V$ and $I$ bands in 2012 and 2014, 
3.62 and 5.59 years after the peak.
In this paper we present the analyses of these 
{\it HST} images to confirm that the star blended with the source is not the planetary host star. 

The paper is organized as follows: Section \ref{sec-lc} presents light curve modeling 
of a slightly different data set than used for the the discovery paper 
due to a re-reduction of MOA light curve data that corrected for systematic errors due to 
differential refraction. Section \ref{sec-radius} determines the source color and angular radius
from the parameters presented in Section \ref{sec-lc}. {\it HST} follow up image analyses 
with different point spread function (PSF) fits are discussed in Section \ref{sec-hst}. 
Subsections \ref{sec-single-psf}, \ref{sec-dual-psf} and \ref{sec-triple-psf} explore the 
results of fitting single star, dual star  and triple star PSFs. The final section, \ref{sec-Conclusion},
presents the upper limit of the lens brightness and proceeds with the calculations of the lens properties.   

\begin{deluxetable}{cccccc}
\tablecaption{Microlensing Model Parameters \label{tab-params}}
\tablewidth{0pt}
\tablehead{\colhead{parameter}&\colhead{units}&\multicolumn{2}{c}{best fit}&\multicolumn{2}{c}{MCMC Averages}\\
&&close&wide&no constraint&lens brightness constrained$^*$}
\startdata
$t_E$& days &10.22&10.29&10.27(0.27)&10.27(0.24)\\
$t_0$&HJD$-2450000$&4656.39&4656.39&4656.39(0.00011)&4656.39(.00011)\\
$u_0$&$10^{-3}$&3.26&3.18&3.21(0.09)&3.21(0.09)\\
$s$ &&0.93&1.08&1.04(0.07)&1.04(0.07)\\
$\theta$&radian&1.95&1.93&1.94(0.02)&1.94(0.02)\\
$q$&$10^{-4}$&3.29&3.49&3.38(0.28)&3.38(0.28) \\
$t_\star$&days&0.054&0.055&0.055(0.00011)&0.055(0.00011)\\
$\chi^2$&&6616.88&6618.93&&\\
\enddata
$^*$ The lens brightness constraint is based the 3-star PSF fits discussed in Sections 
\ref{sec-triple-psf} and \ref{sec-Conclusion}. 
\end{deluxetable}    

\section{Revisiting Light Curve Modeling}
\label{sec-lc}

A single microlens event light curve model uses three non-linear parameters: 
$t_0$ - the time of peak magnification, $u_0$ - the minimum separation between the 
source and the lens in Einstein radius units, and $t_E$ - the Einstein radius crossing time. 
The Einstein radius is given by
$R_E = \sqrt{(4GM/c^2)D_Sx(1-x)}$, where $x = D_L/D_S$ and $D_L$ and $D_S$ are
the lens and source distances, respectively. ($G$ and $c$ are the Gravitational constant
and speed of light, as usual.) 
There are also two linear parameters for each data set: the source flux $f_s$, 
and the blend flux, $f_{bl}$. To fit a binary microlens model, we need three additional 
non-linear parameters: $q$, the lens mass ratio, $s$, the projected separation between the 
lens masses measured in the Einstein radius units, and $\theta$, the angle between the 
source trajectory and the lens axis. Also, binary events often have caustic or cusp 
crossings, which resolve the angular size of the source, so we need to model the finite 
source effects with the source radius crossing time, $t_\star$. In the years since the original
paper on this event \citep{moa310}, we have found that it is possible to improve the photometry
for many events by removing trends due to air mass, differential refraction and seeing that
are observed in the data before and after the microlensing event. The removal of these
systematic error trends can sometimes modify the best fit $t_E$ and source flux, $f_s$,
values, so we thought it prudent to use the new photometry. In this case, the detrended
photometry resulted in slightly different parameters, but no large change in $t_E$ or $f_s$
was seen. We modeled MOA red ($R+I$), CTIO SMARTS $I$, $H$, $\mu$FUN Auckland $R$, 
$\mu$FUN Bronberg unfiltered and PLANET Canopus $I$-band data using the $\chi^2$ 
minimization recipe of \citet{bennettmcmc} to find the best fit wide ($s > 1$) and close ($s < 1$)
models, as shown in Table~\ref{tab-params}. These $s \leftrightarrow 1/s$ degenerate models \citep{griest1998} 
are due to the usual high magnification
separation degeneracy as noted by \citet{moa310}. Once the best fit models were identified,
we ran several Markov Chain Monte Carlos (MCMC) \citep{mcmc} to determine the distribution
of parameters that are consistent with the light curve measurements.
The uncertainties are given by the root mean squares (RMS) variations over the MCMC links
for each parameter, as shown in Table~\ref{tab-params}. The methods of error bar renormalization 
and the calculation of the limb darkening 
effects are similar to those of \citet{moa310}. The difference between the models presented
in Table \ref{tab-params} and those of the discovery paper are not noticeable in a light curve plot. 

\section{Source Radius and Color Determination}
\label{sec-radius}
The models listed in Table \ref{tab-params} yield the source brightness in the CTIO SMARTS 
$I$ and $H$ bands as $I_{\rm CTIO} = 18.93 \pm 0.03$ and $H_{\rm CTIO} = 21.47 \pm 0.03$. 
We use OGLE-III and VVV magnitudes as the calibrated magnitudes in the visible and infra 
red bands respectively.
We calibrate the CTIO $I$ band reference image to the OGLE-III catalog 
\citep{ogle3-phot} using 248 isolated stars of brightness $I_{\rm OGLE_{III}} < 16.0$. 
We obtained the following calibration relation:
\begin{equation}
I_{\rm CTIO} = I_{\rm OGLE_{III}} + 0.057998 (V-I)_{\rm OGLE_{III}} -0.596221 \pm 0.02
\label{eq-I}
\end{equation}
The CTIO $H$ band reference image is matched with the VVV catalog \citep{vvv}
and 139 bright isolated stars are cross identified to obtain the $H$ band calibration relation:
\begin{equation}
H_{\rm CTIO} = H_{\rm 2MASS} + 3.781 \pm 0.004
\label{eq-H}
\end{equation}
The uncertainties in these Equations~\ref{eq-I} and \ref{eq-H} are given by RMS/$\sqrt N$ 
where $N$ represents the number of stars used. Equation~\ref{eq-H} yields a best fit source 
magnitude of $H_{\rm 2MASS} = 17.69 \pm 0.03$.
\begin{figure}
\epsscale{1.0}
\plotone{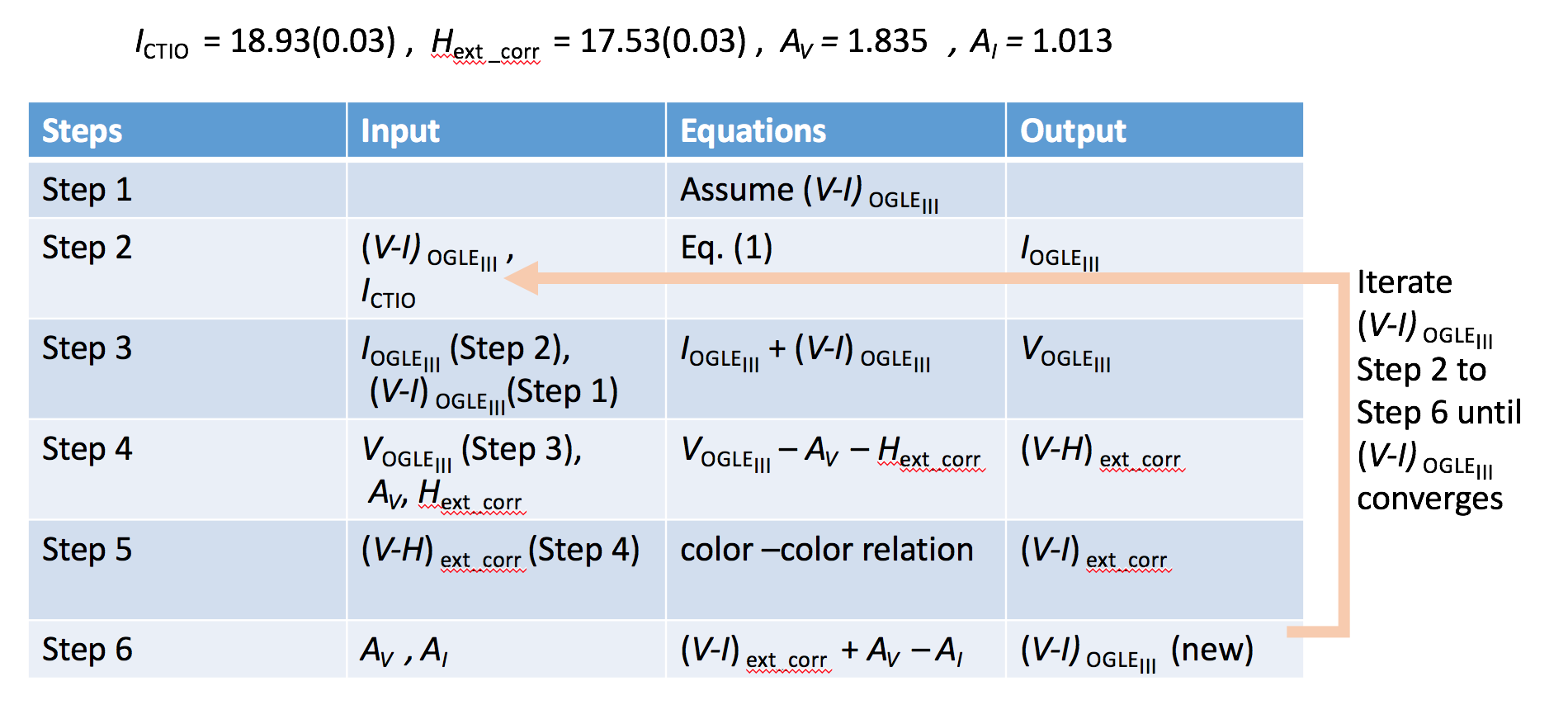}
\caption{The methodology for source color derivation. \label{fig-color}} 
\end{figure}

\begin{deluxetable}{ccccccc}
\tablecaption{Deriving source color using iteration from \citet{kenyon_hartmann} color-color relation
                         \label{tab-color} }
\tablewidth{0pt}
\tablehead{
\colhead{Iteration}&\colhead{$V-I$}&\colhead{$I_{\rm OGLE_{III}}$}&\colhead{$V_{\rm OGLE_{III} }$}&\colhead{$(V-H)_{\rm ext\_corr}$}&\colhead{$(V-I)_{\rm ext\_corr}$}&\colhead{new $(V-I)$} \\\colhead{Number}&
 & Eq~\ref{eq-I} + $I_{\rm CTIO}$ &
 & & &  
}  
\startdata
1 & 1.450 & 19.44 & 20.89 & 1.53 & 0.76 & 1.582 \\
2 & 1.582 & 19.43 & 21.02 & 1.65 & 0.79 & 1.611 \\
3 & 1.611 &19.43 &21.04 &1.68& 0.8 & 1.622  \\
4 & 1.622& 19.43 & 21.05 & 1.69 &0.81&1.631 \\
5 & 1.631 & 19.43 & 21.06 & 1.70 & 0.815 & 1.637  \\
6 & 1.637 & 19.43 & 21.07 & 1.71 & 0.82 & 1.642  \\
7 & 1.642 & 19.43 & 21.07 &1.71  & 0.82& 1.639  \\
8 & 1.639 & 19.43 & 21.07 & 1.71 & 0.82 & 1.640 \\
\enddata

$^*$Refer to figure \ref{fig-color} for details. The color of the star was not readily available in 
Cousins $I$ and Johnson $V$ from observations hence this iteration method was adopted. 
\end{deluxetable}

Our aim is to derive $I_{\rm OGLE_{III}}$ from Equation~\ref{eq-I}, but we do not have
a measurement of $(V-I)_{\rm OGLE_{III}}$. There is only a single CTIO $V$ band measurement 
that is magnified, and that measurement has a 25\% uncertainty. So, we cannot get an accurate
source color measurement from the CTIO $V$ band data. Therefore, we choose an iterative 
method to determine the source color with the help of the color-color relations of
\citet{kenyon_hartmann}. This iterative method utilizes $I_{\rm CTIO}$, $H_{\rm CTIO}$ and 
Equations~\ref{eq-I} and~\ref{eq-H}. The steps of this method are described in detail below and in
Figure~\ref{fig-color}. We use the extinction values of $E(V-I) = 0.822$,  $A_I  = 1.013$
and $A_V = 1.835$ from \citet{nataf13}. From the knowledge of $A_I$ and $A_V$, 
using \citet{cardelli} we find $A_H = 0.16$. The extinction corrected $H$ band source magnitude is 
$H_{\rm 0} = H_{\rm 2MASS} - A_H = 17.53 \pm 0.03$. Also we use the equation~\ref{eq-H} providing 
the source magnitude $H_{\rm 2MASS} = 17.69 \pm 0.03$ and the Equation 1 for this iterative method.\hfil\break
{\bf Step 1.} We assume $(V-I)_{\rm OGLE_{III}} = 1.45$.\\ 
{\bf Step 2.} With the value of $I_{\rm CTIO}$ and $(V-I)_{\rm OGLE_{III}}$, we determine 
    $I_{\rm OGLE_{III}}$ from equation~\ref{eq-I}.\\ 
{\bf Step 3.} With $(V-I)_{\rm OGLE_{III}}$ from step 1 and $I_{\rm OGLE_{III}}$ from step 2,
  we determine $V_{\rm OGLE_{III}} = I_{\rm OGLE_{III}} + (V-I)_{\rm OGLE_{III}}$.\\ 
{\bf Step 4.} From $V_{\rm OGLE_{III}}$ in step 3, $A_V = 1.835$ and $H_{\rm 0} = 17.53 \pm 0.03$, 
we find the extinction corrected $(V - H)_0$.\\
{\bf Step 5.} This extinction corrected $(V - H)_0$ gives a new value of the extinction corrected  
$(V-I)_0$ using the color -color relation of \citet{kenyon_hartmann}.\\
{\bf Step 6.} We add  $E(V-I)$ to this value to get the new $(V-I)$ for our next iteration.
  
In the next iteration we use this new $(V-I)$ and repeat the steps  2 to 6 and we continue this 
iteration over $(V-I)$ until its value converges. The method is shown in Figure \ref{fig-color}
and the values for each iteration round are shown in Table~\ref{tab-color}.
This method converges to the source color $(V-I)_{\rm OGLE_{III}} = 1.64~\pm~0.06$ 
and the source magnitude 
$I_{\rm OGLE_{III}} = 19.43~ \pm~ 0.04, V_{\rm OGLE_{III}} = 21.07~\pm ~0.07$. 
With the extinction values given above, we find an extinction corrected magnitude of
$I_{s0} = 18.41 \pm 0.04$ and an extinction corrected color of $(V-I)_{s0} = 0.82 \pm 0.06$.
This is somewhat redder than the $(V-I)_{s0} = 0.69$ claimed by the discover paper
\citep{moa310}, but is is consistent with the estimate of $(V-I)_{s0} = 0.75 \pm 0.05$ based
on a spectrum measured at high magnification \citep{bensby13}. Using the source color from \citet{bensby13} we obtain $I$ and $V$ band magnitudes of source are $19.44 \pm 0.04$ and $21.01 \pm 0.07$ respectively. These values are consistent with the values obtained in our iteration method. 
According to \citet{kenyon_hartmann} the source star is most likely a G5-K0 star. 
   
From the dereddened source magnitude and color, we obtain the source radius using 
the relations from \citet{Boyajian2014}. From the fit parameters, $t_\star = 0.055$ days 
and $t_E$ = 10.27 days and from the knowledge of $\theta_\star$, $\theta_E$ is 
calculated from $\theta_E = \frac{\theta_\star}{t_\star}\times{t_E}$ and 
$\mu_{\rm rel} = \frac {\theta_\star}{t_\star} = \frac{\theta_E}{t_E}$.   
\begin{gather}
\log_{10}(2\theta_\star) = 0.53665 + 0.072703 (V-H) -0.2 H \nonumber\\
\theta_\star = 0.719 ~\pm~ 0.023 {~\rm \mu as}
\end{gather}
\begin{gather}
\log_{10}(2\theta_\star) = 0.53026 +0.16595 (I-H) -0.2 H \nonumber\\
\theta_\star = 0.743~ \pm~ 0.054 {~\rm \mu as}
\end{gather}
The discovery paper has $\theta_\star = 0.76 ~\pm~ 0.05 {~\rm \mu as}$ and 
$\mu_{\rm rel} = 5.1~\pm~ 0.3$ mas/yr. The relative proper motion $\mu_{\rm rel}$ mentioned 
in this section is the geocentric relative proper motion. We derive the weighted average of $\theta_\star = 0.723 \pm 0.021 ~ \mu$as. The corresponding  
$\mu_{\rm rel} = 4.85~\pm~0.15$ mas/yr. The uncertainty is calculated from the standard error over MCMC chains. The uncertainty arises from the 
$0.06\,$mag uncertainty in the source color, the $0.06\,$mag uncertainty in the source 
magnitude. There is also $2.4\%$ and $7\%$ uncertainties 
in the source star angular diameter arising from the scatter about the source star size-color relations in Equations 3 and 4 from \citet{Boyajian2014}. Equation 3 is not included in \citet{Boyajian2014} paper, it is obtained through a private communication with T. S. Boyajian. She reports a $2\%$ scatter of measurements around the fit in $V-H, H$ relation. \citet{kervella_dwarf}  reports a $1.12\%$ uncertainty in the $\theta_*$ value for their $V-H,H$ relation. The difference is that
\citet{Boyajian2014} just reports the scatter in the fit, while \citet{kervella_dwarf} subtracts the
scatter due to the photometric measurement uncertainties. So, the \citet{Boyajian2014}
method is very conservative, and the \citet{kervella_dwarf}’s method should be more
accurate as long as the photometry error bars are accurately estimated.
If the photometry error bars are overestimated, then the error bars on the
relation will be underestimated. Because of this controversy, we prefer to use the more conservative approach of \citet{Boyajian2014}. Also there is uncertainty due to 
$t_\star$ and $t_E$ (see Table \ref{tab-params}). 
%

\section{{\it HST} Image Analyses}
\label{sec-hst}
The event MOA-2008-BLG-310 was observed with the {\it HST} Wide Field Camera 3--Ultraviolet Visible 
(WFC3-UVIS) instrument on February 22, 2012  as part of the program GO 12541 with a second
epoch of observations on February 09, 2014. Both observation epochs used both the
F814W and F555W passbands (which are {\it HST} versions of $I$ and $V$ bands). 
In each epoch, eight images were taken with the exposure times of 70 and 125 sec 
for F814W and F555W filters, respectively. (Hereafter, we refer to the F814W and F555W passbands
as the {\it HST} $I$ and $V$ bands.)
To obtain these many short dithered exposures, it
was necessary to read out only a small 1k$\times$1k subset of each image. Each WFC3-UVIS
pixel subtends approximately $40\,$mas on a side. These dithered images were reduced and 
stacked following the methods described in \citet{andking00,andking04}. 
Fifteen bright isolated stars  with $1.0 \leq (V-I)_{\rm OGLE_{III}} \leq 2.1$ and $I < 17.0$ are cross 
identified and matched between {\it HST} stack images and stars from the OGLE-III catalog
\citep{ogle3-phot} to obtain the following calibration relations for the 2012 epoch:
\begin{eqnarray}
I_{\rm OGLE_{III}} = 29.078375 + I_{HST} + 0.004528 (V-I)_{HST} \pm 0.02023 \nonumber\\
V_{\rm OGLE_{III}} = 30.593088 + V_{HST} - 0.065837 (V-I)_{HST} \pm 0.03233
\label{eq-hst2012}
\end{eqnarray}
Similarly for the 2014 epoch, fourteen stars with $1.0 \leq (V-I)_{\rm OGLE_{III}} \leq 1.9$ 
and $I < 17.0$ are matched to obtain these calibration relations:
\begin{eqnarray}
I_{\rm OGLE_{III}} = 29.054552 + I_{HST} - 0.024727(V-I)_{HST} \pm 0.023431 \nonumber\\
V_{\rm OGLE_{III}} = 30.563614 + V_{HST} - 0.092588 (V-I)_{HST} \pm 0.043662
\label{eq-hst2014}
\end{eqnarray}
The uncertainties are the standard errors of the mean. The uncertainty due to PSF fitting is also included.

The images in each band are reduced with correction for CTE (Charge Transfer Efficiency) 
losses using a method developed specifically for WFC3/UVIS based on the algorithm described in \citet{anderson2010} for ACS. The images are taken in a custom dither pattern that ensured the most uniform possible pixel-phase coverage for the eight exposures. We adopted the first exposure of each passbands as the reference image. The we measured the stars in this image and in all the other images with a library PSF and corrected them for distortion. We then used the positions of the stars common to each exposure and the reference exposure to define a 6-parameter linear transformation from each frame into the reference frame. This allowed us to transform the location of each pixel into the reference frame, both for the purposes of stacking and for the purposes of using them as simultaneous constraints in modeling the scene. We stack all these individual images 
into a single frame \citep{drizzle}, or stack image, for each passband and identify
the target object in the stack images using the NACO VLT  high resolution image presented in 
the discovery paper \citep{moa310}, which was identified based on the location found in difference
images taken near peak magnification. Next we select about 23 isolated stars with a color within 
a 150 pixel radius of the target object that have a color similar to the target.
We use these stars to do another round of mapping between different frames. The goal of this 
step is to generate more precise local coordinate transformations from each frame to the 
reference frame and to derive a more accurate PSF for the target. We then
use these transformations to extract the pixels in the vicinity of the target from each exposure 
and transform their locations into the reference frame and then solve for the effective PSF
\citep{andking00} appropriate for the target object.
 
The main problem in dealing with HST images relative to ground based images is that HST images 
are undersampled. This means a large portion of the flux of a star falls inside one pixel. As a 
result, several different PSF models can provide equally good fits to the data \citep{andking00}. The pixels are too wide to sample all of the information that the telescope is delivering to the detector.
This is not a big problem for determining total flux using aperture photometry as long stellar images
are reasonably isolated. But it is a big problem for astrometry. The degeneracy in the PSF fits yields 
degeneracies in the positions of stars, making it imposible to measure precise stellar positions. 
PSF fitting photometry routines designed for ground-based data, like DAOPHOT \citep{Daophot} 
and DOHPOT \citep{dophot} are not designed to deal with undersampled data, and they are
ill-equipped to deal with intrapixel scale detector sensitivity variations, which can be important
for undersampled data. Theoretically, one might attempt to deal with undersampling by separately
determining the instrumental PSF (iPSF) and the sub-pixel scale response function, but 
\citet{andking00} point out that it is simpler and more accurate to deal with the effective PSF (ePSF),
which is the convolution of the iPSF and the detector response,
\begin{equation}
\psi_E = \Re \times \psi_\imath \ ,
\end{equation} 
where $\Re$ and $\psi_\imath$ are intrapixel sensitivity function and iPSF respectively. 
We selected about 19 isolated stars within 150 pixel radius and within 0.1 magnitude color and 
0.5 magnitude $V$ band brightness of the target star. These stars were used to build the ePSF. We used all the individual images of these
stars to determine the effective PSF following the method of \citet{andking00,andking04}. 
We select a pixel region centered on each of the PSF-contributing stars and divide it into a grid with a quarter pixel interval.  
The ePSF is evaluated on these grid points, and the intermediate points are interpolated using 
cubic spline interpolation. The value of the ePSF is the fraction of a star's light that should fall 
in a pixel centered on the specified coordinates. Hence:
\begin{equation}
 P_{ij} = z_*\psi_E(i-x_*,j-y_*) + s_*
 \label{eq-Pij}
\end{equation} 
 or conversely,
\begin{equation}
\psi_E (i-x_*,j-y_*) = ( P_{ij} -s_*)/z_*
\label{eq-psiE}
\end{equation} 
where  ${x_*,y_*}$ are the position of the star and $P_{ij}$ is the observed pixel value 
of the pixel centered at $(i,j)$. The parameters $s_*$ and $z_*$ are the background flux and 
the the total flux of the star, respectively. We start with the total flux of the star measured 
by aperture photometry  as $z_*$ and the centroid of the star as the position ${x_*,y_*}$. For 
the background flux we calculated the flux between 8.5 and 13 pixels from the center of the star. 
The ePSF is computed with an iterative procedure. In each iteration, 
we determine the ePSF for each of those 19 stars in each individual image. Then we average 
the ePSFs of each star from all the images combined. It is this step of averaging of the 
ePSFs from all the dithered images that helps to overcome the undersampling problem. Now 
we use this average ePSF to get a best fit of the star by minimizing:
\begin{equation}
 \chi2 = \sum_{i,j} w_{ij}[P_{ij} -s_* -z_*\psi_E(i-x_*,j-y_*)]^2
\end{equation}
We do not fix the value of (${x_*,y_*}$) in this minimization procedure, 
so it yields a new set of values for star positions which is used to calculate the effective 
PSF from Equation~\ref{eq-Pij} in the next iteration step. This iteration procedure converged
to our final ePSF after 6 iterations.
 
Once we have determined our ePSF model, we are ready for fit our target object with
two star models, so that both the source and lens stars can be included. We also consider three
star models in cases (like this one) where the two-star models don't provide a good fit to the
properties of the source and lens stars. Our model for the total star flux distribution changes from 
$z_*\psi_E(i-x_*,j-y_*)$ for a single star to $f_1\psi_E(i-x_1,j-y_1) + (1-f_1)\psi_E(i-x_2,j-y_2)$ 
for the dual star model. This increases the number of model parameters by 3, for the 
brightness and coordinates of the second star. The parameter $f_1$ denotes the ratio
of the star-1 brightness to the total stellar brightness of both stars, so
$(1-f_1)$ is the fractional brightness of the second star. The parameters
${x_1,y_1}$ and ${x_2,y_2}$ are the positions of the two stars. We have two different 
strategies for these ePSF fits. For the simplest, two-star
models, we start with a simple grid search that gives us a list of $\chi^2$ values for
all parameter sets that fall on the parameter grid. This is robust, but inefficient. It is much
more efficient to use the Markov Chain Monte Carlo (MCMC) method, which avoids 
highly unlikely parameter choices. We use the MCMC method to fit $x_1,y_1,x_2,y_2,f_1$ in order to 
minimize the following $\chi^2$ for each individual image
\begin{equation}
\chi2 = \sum_{i,j} w_{ij}[P_{ij} -s_* - f_1\psi_E(i-x_1,j-y_1) + (1-f_1)\psi_E(i-x_2,j-y_2)]^2 \ .
\end{equation}
The ``chain" of solutions generated with the MCMC serves as a probability distribution of
parameters that we use to determine the uncertainties on the PSF fit parameters.
    
The RA and Dec of the stack images are related to x and y positions using the 
Equations~\ref{eq-2012} and \ref{eq-2014} for 2012 and 2014 data respectively:
 \begin{gather}
 \label{eq-2012}
{\rm RA} = 2.56\times10^{-2}(x-558.4) +3.08\times10^{-2}(y-623.2)+966819.47\\
 {\rm Dec} = 3.05\times10^{-2}(x-558.4) - 2.56\times10^{-2}(y-623.2)-125199.89\nonumber
  \end{gather}
 \begin{gather}
  \label{eq-2014}
{\rm RA} = 2.53\times10^{-2}(x-498.6) +3.09\times10^{-2}(y-635.1)+966818.03\\
 {\rm Dec} = 2.95\times10^{-2}(x-498.6) - 2.38\times10^{-2}(y-623.2)-125201.95\nonumber
 \end{gather}
The RA and Dec are expressed in arcseconds. From these relations it is clear that 
the $x$ and $y$ pixel positions of lens and source are different in different frames. But the 
relative separation between the lens and the source is independent of the frames. 
Each pixel is $\sim$40 mas. We will be using separations in x and y pixel 
coordinates in this paper. 

\subsection{Single star PSF Fit}
\label{sec-single-psf}
The first step in our PSF modeling is to do single star fits to all the stars in the frame. Such
fits are used for the calibration of the {\it HST} photometry presented in equations~\ref{eq-hst2012}
and \ref{eq-hst2014}. Aside from the calibration, we are primarily interested in the ``target" star,
which is at the position of the source star (MOA-2008-BLG-310S). Since the ePSF model is
fixed from a fit to a set of stars with color similar to the source star, there are on three
parameters that describe the ePSF fit to the target star. These are 
the pixel position of the star in x and y and the total flux. In the 2012 epoch, the single star 
PSF fit at the position of the target yields calibrated magnitudes of $I = 19.29 \pm 0.02$
and $V = 20.81 \pm 0.03$. Similarly, the 2014 epoch images yield target magnitudes of
$I = 19.27 \pm 0.02$ and $V = 20.80 \pm 0.04$. These magnitude uncertainties come from 
the PSF fit uncertainties and the calibration uncertainties. It is clear from these magnitudes that the 
target object is brighter than the source, $I_{S} = 19.43 \pm 0.05, V_{S} = 21.07 \pm 0.07$, and this
implies that there is at least one additional star blended with the source star, as found by
\citet{moa310} with AO images from the VLT. However, due to the stability of the {\it HST}
PSF, we can use our {\it HST} images to do photometry and astrometry of the stars that 
contribute to the target, even though these stars remain unresolved. The next step is to fit
the target with a dual star model to see if the target can be explained as a combination of
the source and lens stars.

\subsection{Dual Star PSF Fits}
\label{sec-dual-psf}
In dual star PSF fits, we expect to fit and detect the source and the lens. The source magnitudes in $I$ and $V$ bands are $19.43 \pm 0.04$ and $21.07 \pm 0.07$ respectively. The geocentric lens-source relative proper motion is $4.81 \pm 0.15$ mas/yr. Since the light curve data of this event was very well covered, there is a very small scope of the light curve model to be incorrect. Following the relative proper motion, we expect to see a lens- source separation of about 17.4 $\pm$ 0.4 and 27.4 $\pm$ 0.7 mas in the first and the second epoch respectively. If we can detect the lens and the source, the brightness of one star should be consistent with the brightness of the source in each passband and the separation measured between the two stars should be consistent with the predicted lens-source separation in each epoch. 
\subsubsection{Unconstrained Best Fit}
\label{sec-dual-psf_unconst}

For the dual star ePSF models there are total of 6 parameters to fit. These are the pixel positions of the 
two stars in $x$ and $y$, and the total flux and the flux ratio between the two stars. The dual star
models were run with both the grid search and MCMC methods, which yielded essentially
identical results. These are presented in the top section of Table~\ref{tab-dual-fit}. Note that
the $\chi^2$ values for each fit were initially somewhat larger than the values reported in this
Table, which were initially estimated on the basis of Poisson noise and read-out noise in the 
individual {\it HST} images. However, it is reasonable to presume that there is an additional
uncertainty due to imperfections in the ePSF models. Therefore, we renormalize the uncertainties
to give $\chi^2/{\rm dof} = 1$ for each passband. The number of pixels fitted in 
the 2014 $I$ and $V$ bands are 207 and 202, and the correction factors for 
the 2014 $I$ and $V$ bands are 1.73 and 1.53 respectively. Similarly the number of pixels 
fitted in 2012 $I$ and $V$ bands are 200 and 199, while the correction factors are 1.42 and 1.27,
respectively. This is similar to the procedure applied to light curve modeling for
virtually all planetary microlensing events \citep{bennett08}.

These solutions and their uncertainties show that in both the epochs, neither stars' brightness 
matches the source magnitudes of $I_{S} = 19.43\pm 0.04$ and $V_{S} = 21.07 \pm 0.07$.
The source is brighter than star 1 by 2.6-$\sigma$ and 2.8-$\sigma$ in the $I$ band and 
by 1.6-$\sigma$ and 1.7-$\sigma$ in the two epochs of $V$ band data, so the total
significance of the difference between the star 1 flux and the source flux is about 4.2-$\sigma$
when all 4 measurements are considered.
Thus, these solutions are not consistent with the microlensing light curve data, which requires
a source brighter than star 1. We attribute this difference to minor problems with the PSF model,
which are overcome in the next section, which describes source flux constrained fits.
\begin{landscape}
\begin{deluxetable}{ccccccccc}
\tablecaption{List of Dual Star Fits \label{tab-dual-fit}}
\tablewidth{0pt}
\tablehead{\colhead{Dual Star Fit}&\colhead{Year}&\colhead{Filter}&\multicolumn{2}{c}{Magnitude$^*$}&\colhead{Separation}&\multicolumn{2}{c}{Separation star 2 - star 1}&\colhead{$\chi^2$}\\ 
& & & \colhead{Star 1}&\colhead{Star 2}&(mas)&\colhead{$\vartriangle$x}&\colhead{$\vartriangle$y}&
}
\startdata
\multirow{4}{*}{Best Fit}&\multirow{2}{*}{2012} &$I$ &19.84(0.15) & 20.31(0.24) & 14.1(3.2)&9.5(2.6)&10.1(2.6)&194.3\\
& &$V$ &21.37(0.18) & 21.74(0.56) & 15.2(2.9)&6.7(2.5)&13.1(2.1)&193.1\\
&\multirow{2}{*}{2014}&$I$ &19.86(0.14) & 20.28(0.18) & 12.2 (3.3)&10.6(1.8)&4.7(2.8)& 201.1\\
& &$V$& 21.47(0.22)& 21.64(0.27)&11.6(3.7) &10.3(2.2)&6.2(3.1)& 195.9 \\ \\
\multirow{4}{3 cm}{\centering{Source Flux Constrained}}&\multirow{2}{*}{2012} &$I$ &19.47(0.05)&21.35(0.29)&16.6(2.1)&11.2(1.2)&12.4(1.6)&204.2\\
& &$V$ &21.11(0.12)&22.26(0.38)&16.1(2.9)&9.2(2.1)&11.6(2.1)&199.2\\
&\multirow{2}{*}{2014}&$I$&19.45(0.05)&21.43(0.31)&14.1(2.1)&12.4(1.2)&6.1(1.2)&210.9\\
& &$V$&21.11(0.11)&22.31(0.43)&13.5(2.4)&11.6(1.2)&7.6(2.1)&200.9 \\\\
\multirow{4}{2.5 cm}{\centering{Source Flux and Separation Constrained}}&\multirow{2}{*}{2012} &$I$&19.46(0.04)&21.41(0.21)&17.4(3.4) &11.4(2.5)&12.7(2.3)&214.5\\
& &$V$ &21.09(0.13)&22.28(0.28)&17.3(3.6)&6.5(2.8)&13.6(2.2)&206.2 \\
&\multirow{2}{*}{2014}&$I$ &19.38(0.06)&22.18(0.27) &26.9(7.7) &20.3(7.1)&14.5(3.1)&233.9\\
& &$V$&21.02(0.09)&22.66(0.48)&26.8(9.3)&23.8(8.3)&12.5(4.2)&218.7\\
\enddata
\\This Table presents magnitudes calibrated to the OGLE-III scale.  
\end{deluxetable}
\end{landscape}

  \subsubsection{Source Flux Constrained Fits}
  
To find a solution consistent with the source magnitude from our light curve models, 
we add a constraint on the flux of the source star in our MCMC chains. The constraint is imposed
by adding a term of the form  $\exp [(f_1-f_s)^2/(2\sigma^2_{f_s})]$ to each model $\chi^2$ 
calculation. As discussed in Section~\ref{sec-lc}, the parameter $f_s$ represents the
source flux and $\sigma_{f_s}$ is its uncertainty. This uncertainty comes from the 
light curve models, the calibration relations \ref{eq-I}, \ref{eq-hst2012}, \ref{eq-hst2014}
and the source star color determination described in Section~\ref{sec-radius}.
The parameter $f_1$ represents the source flux from the dual star PSF model,
because we identify star 1 with the source. Our constraint term forces the source brightness
to be consistent with the light curve models.
The results of these constrained MCMC runs are shown in Table \ref{tab-dual-fit}. 
The source brightness constraint raises $\chi^2$ by an amount ranging from 5.0 to 9.9. These
values are a bit above what one would expect from Gaussian statistics, but we expect that this
is due to inadequacies in the PSF model.

The predicted lens-source separations are 17.4 $\pm$ 0.4 mas and 27.4 $\pm$ 0.7 mas in 2012 
and 2014, respectively.
In the 2012 epoch, 3.62 years after peak magnification, the separation between the two stars is 
$\sim 1$-$\sigma$ away from the predicted 
lens-source separation. In the 2014 epoch, 5.59 years after peak magnification,
this separation is at least 7$\sigma$ away from the predicted value. This is not
consistent with the assumption that the excess flux is due to the lens star.

  \subsubsection{Source Flux and Lens - Source Separation Constrained Best Fit} 
  \label{sec-flux_sep_const}
  
The results of the source flux constrained fit show that the separation between the source 
and the blend star is not consistent with the predicted lens-source separation. Hence, the excess
flux blended with the source is not dominated by the lens star.  As a further check on this 
blend = lens star model, we have performed fits with both the source flux and the lens--source
separation fixed.
  
For the source flux and separation constrained fit, we add the extra term 
$\exp [-(f_1-f_s)^2/(2\sigma^2_{f_s}) - (s_{12} - s_{\rm lc})^2/(2\sigma_{s_{\rm lc}}^2)]$ 
to the $\chi^2$ for each link in the MCMC. The parameters $f_1$, $f_s$, and $\sigma_{f_s}$ 
refer to the flux of star 1, the source flux, and the uncertainty in the source flux from the
light curve models, while the parameters $s_{12}$, $s_{\rm lc}$, and $\sigma_{s_{\rm lc}}$
refer to the star 1--2 separation, the lens--source separation predicted by the light curve model, and 
the uncertainty in the light curve model separation prediction.
The first term in the exponential is the source flux constraint, as mentioned previously. 
The second term in the exponential similarly is the lens--source separation constraint.
This term forces the fit separation between stars 1 and 2 in the image fits to match, 
within the measurement uncertainty, the light curve model prediction of the lens--source
relative proper motion, $\mu_{\rm rel}$. The lens--source separation, $s_{\rm lc}$ term 
is determined from the light curve model by $s_{\rm lc} = \mu_{\rm rel} \Delta t$, where 
$\Delta t$ is the time interval between the microlensing event peak and the time of the
{\it HST} observations. We should noted that the separation measured between the 
two stars in $HST$ frame, $s_{12}$, is directly related to the relative proper motion, 
$\mu_{\rm rel}$ in the heliocentric frame, while the $\mu_{\rm rel}$ value determined 
from the light curve model, described in Section \ref{sec-radius} determined in a
``geocentric" reference frame that that moves with a constant velocity that matches the
Earth's velocity at the peak of the microlensing event. The relation between the relative
proper motion in these two reference frames is given by the Equation  \citep{dong-moa400}
\begin{equation}
\bm{\mu}_{\rm rel,H} = \bm{\mu}_{\rm rel,G} + \frac{{\bm v}_{\oplus} \pi_{\rm rel}}{\rm AU} \ ,
\label{eq-mu_helio}
\end{equation}
where ${\bm v}_{\oplus}$ is the projected velocity of the earth relative to the sun at the time of peak 
magnification.  The relative parallax $\pi_{\rm rel} \equiv 1/D_L - 1/D_S$ is related to the 
lens mass by the following relation,
\begin{equation}
\pi_{\rm rel}  =  {c^2\over 4G} \theta_E^2 {{\rm AU}\over M_L} \ ,
\end{equation}
where $D_L$ and $D_S$ are the distances to the lens and source, respectively. This implies that 
\begin{equation}
\bm{\mu}_{\rm rel,H} = \bm{\mu}_{\rm rel,G} + \frac{{\bm v}_{\oplus} c^2 \theta_E^2 }{4G M_L} \ .
\label{eq-mu_helioM}
\end{equation}
From Equation~\ref{eq-mu_helio}, we see that the difference between the Heliocentic
and Geocentric proper motions is minimized when the relative parallax, $\pi_{\rm rel}$, is
small, i.e.\ when $D_L \approx D_S$. From equation~\ref{eq-mu_helioM}, we can also
see that corresponds to a large lens mass. Due to the relatively small angular Einstein
radius, $\theta_E$, the light curve model for this event predicts a low lens mass, unless
the lens is quite close to the source. This implies that if the lens star dominates the blend
flux, then we have a large lens mass, $M_L$, and small relative proper motion, $\pi_{\rm rel}$.
This provides us an easy way to deal with the fact that the transformation to Heliocentric
coordinates, equation~\ref{eq-mu_helio}, depends on the direction of the lens--source
relative motion. We simply assume the lens mass derived from the lens = blend assumption
in the discovery paper \citep{moa310}, and then include the direction uncertainty as a
contribution to the uncertainty in $|\bm{\mu}_{\rm rel,H}|$. This yields 
$|\bm{\mu}_{\rm rel,H}| = 4.98 ~\pm~ 0.31\,$mas/yr. This value of $|\bm{\mu}_{\rm rel,H}|$ is used 
to constrain the star 1--2 separation in the source flux plus lens--source separation constrained
fit. 

The results of these fits are summarized in Table~\ref{tab-dual-fit}. The uncertainties for the
values in this table are the root square of the distributions in the MCMC. For the magnitudes,
we also include the uncertainties in the calibration relations. The results in this table indicates that the
2014 epoch fits show a significant increase in $\chi^2$, when the lens--source separation constraint
is added, with increases of $\Delta\chi^2 = 23.0$ and $\Delta\chi^2 = 17.8$ in the $I$ and $V$
bands, respectively. The increases in the $\chi^2$ values for the 2012 data are smaller
($\Delta\chi^2 = 10.3$ and $\Delta\chi^2 = 7.0$ for $I$ and $V$, respectively), as expected
because the lens--source separation was smaller in 2012. Thus, we conclude that the 
blend flux is not dominated by flux from the lens star.


\subsection{Triple star PSF Fit for the Source, Lens and an Additional Star}
\label{sec-triple-psf}

In the last section, we showed that the extra flux on top of the source is not due dominated
by the lens star. This implies that there must be an additional star blended with the source that
must contribute significant flux. If the lens star is very faint, then the source--flux constrained
fits may accurately describe the {\it HST} data. But, in the more general case, there should
be a total of three stars: the source, the lens, and another star that contributes most of the 
excess flux. 

\begin{figure}
\epsscale{1.0}
\plotone{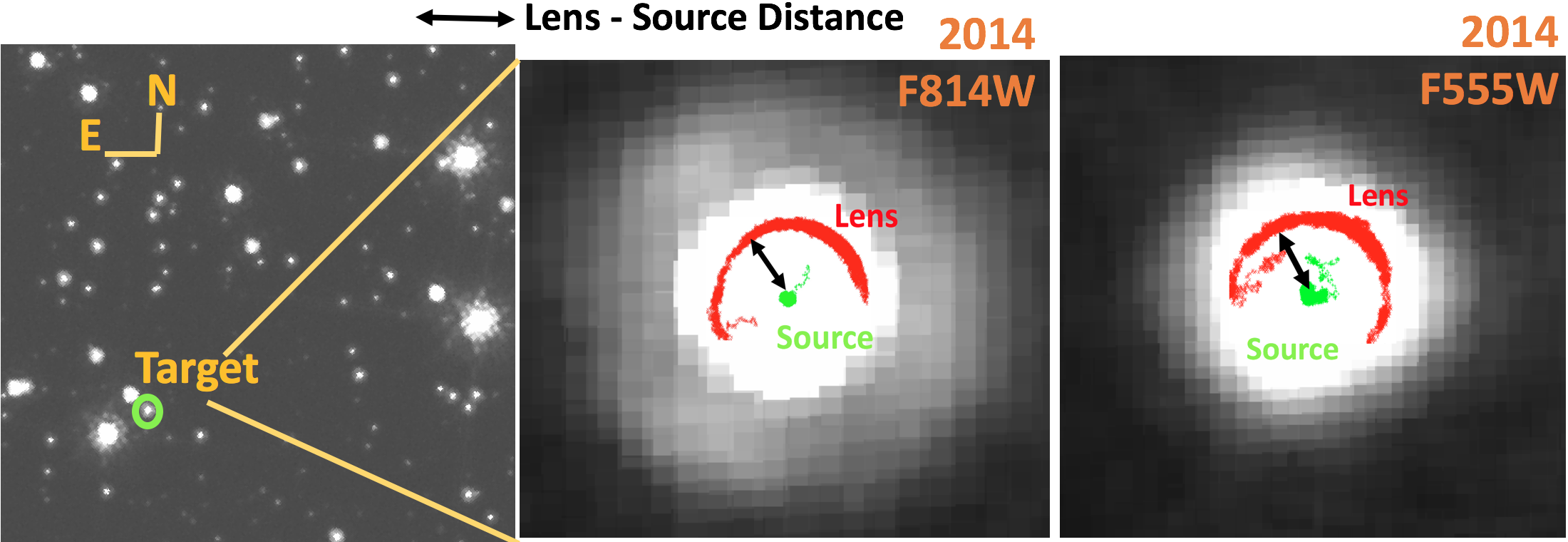}
\caption{Left: The 2014 F814W ($I$-band) stack image showing the target object. 
Middle: The $100\times$ super-sampled stack image showing the source and the lens positions from the 
3-star $I$-band MCMC. Right:  The $100\times$ super-sampled image showing the source and lens 
positions from the 3-star F555W ($V$-band) MCMC.  In both of these 2014 MCMC runs, the source
flux and lens--source distance were constrained, but the direction of the lens--source vector
was not constrained.. This separation conditions are clear from the lens and source positions 
presented here. The source positions are confined to the center, while the lens positions largely
follow an arc.} 
\label{fig-triple}
\end{figure}

\begin{figure}
\epsscale{1.0}
\plotone{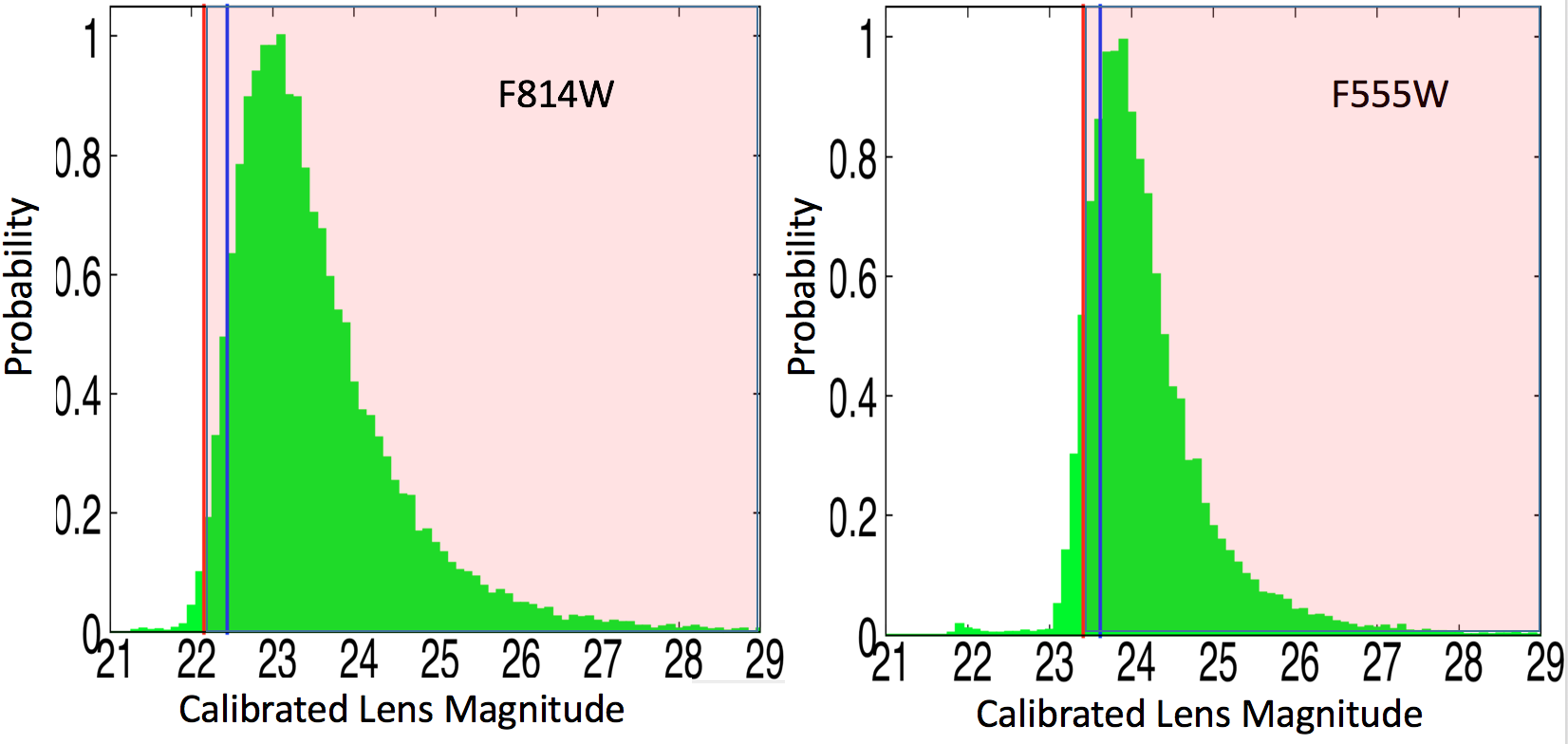}
\caption{The lens flux distribution for the 3-star PSF fits with constraints on both the source flux and 
the lens--source separation for the 2014 F814W and F555W images on the left and right, respectively. 
The red and the blue lines mark the 99\% and 95\% confidence level upper limits on the lens brightness. 
The red shaded regions denote the lens magnitudes fainter than the 99\% c.\ l.\  
upper limit of the lens brightness. These upper limit on the lens brightness to used to 
determine the upper limit on the lens mass presented in Table \ref{tab-params-histogram-meas}.
\label{fig-histogram0}}
\end{figure}

In our triple star PSF fits, the source flux and the lens--source separation are constrained as in
the dual star fits discussed in Section~\ref{sec-flux_sep_const}, but we now add an additional
star. This additional star is presumably either a companion to the source or the lens, or 
else a nearby star. There is no constraint on the position of this star. The direction of the
lens--source separation vector is also unconstrained. For this fit, three additional parameters 
were introduced: the two position coordinates for this third star and the flux ratio between this 
third star and the lens star. We maintain the same error bar normalization as discussed in
Section~\ref{sec-dual-psf_unconst}. The results of these fits are shown in Table \ref{tab-triple}. 
The lens and the source positions from these MCMC runs are shown in Figure \ref{fig-triple}. 
The source positions form a clump in the center for both the $I$ and $V$-band fits, while the
lens positions form an arc around it as shown in the Figure \ref{fig-triple}.  
We calculated the calibrated lens flux for each link in the MCMC chains. The lens flux distribution 
for the 2014 images in the $I$ and $V$ passbands are shown in Figure~\ref{fig-histogram0}. 
The 2012 images show a lens flux distribution consistent with the 2014 results. 
The parameters from this fit are presented in Table \ref{tab-triple}. We use
polar coordinates to describe the lens--source separation, since Figure~\ref{fig-triple} 
shows that the lens positions are largely distributed in an arc. The uncertainties in the lens--source 
separation direction and the lens and additional star's brightness are high, largely because
the data do not demand any light from the lens star. That is, the source flux constrained 2-star fits
shown in Table~\ref{tab-dual-fit} provide an acceptable fit to the data, so the brightness and
position of the lens star will cannot be well constrained. The uncertainty in the lens star brightness
increases the uncertainty in the brightness of the additional blend star.
About $\sim$40 and $\sim$100 source and lens positions fall outside the central clump and arc
distributions for the F814W and F555W images, respectively. The total number of links in the
Markov chains used to create these figures were $\sim 29,000$ and $\sim 18,500$, for
F814W and F555W, respectively, so these represent $\simlt 0.5$\% of the distribution.
These constrained fits are used in Section~\ref{sec-upper} to derive the upper limit of brightness on 
lens and planetary host star.

\subsubsection{Upper Limit Calculation on the Lens Brightness and Mass}
\label{sec-upper}

Figure~\ref{fig-histogram0} shows the distribution of the $I$ and $V$-band lens star
magnitudes from the 2014 source flux and lens--source separation constrained triple star 
fits discussed in Section~\ref{sec-triple-psf}. We use these histograms to determine the upper 
limit on the lens brightness. In 2014 $I$ band and $V$ band, 99$\%$ of the lens magnitude 
distribution lies fainter than $I > 22.15$ and $V > 23.41$. The corresponding limits from the 2012 data
are $I > 21.27$ and $V > 22.02$. These are weaker due to the smaller lens--source separation
in 2012.

The microlensing light curve model provides the mass-distance relation,
\begin{equation}
M_L = {c^2\over 4G} \theta_E^2 {D_S D_L\over D_S - D_L} 
       =  {c^2\over 4G} \theta_E^2 {{\rm AU}\over \pi_{\rm rel}}
       = 0.9823\,\msun \left({\theta_E\over 1\,{\rm mas}}\right)^2\left({x\over 1-x}\right)
       \left({D_S\over 8\,{\rm kpc}}\right) \ ,
\label{eq-m_thetaE}
\end{equation}
where $x = D_L / D_S$. This relation can be combined with a mass-luminosity relation 
to obtain the mass and the distance of the host star and the planet following the 
method in \citet{ogle169}. We use the empirical mass-luminsity
relations of \citet{henry1993}, \citet{henry1999} and \citet{delfosse00}.
For $M_L > 0.66\,\msun$, we use the \citet{henry1993} relation; for
$0.12\,\msun < M_L < 0.54\,\msun$, we use the \citet{delfosse00} relation; and for
$0.07 \,\msun < M_L < 0.10\,\msun$, we use the \citet{henry1999} relation. In between these
mass ranges, we linearly interpolate between the two relations used on the
boundaries. We interpolate between the \citet{henry1993} and the \citet{delfosse00}
relations for $0.54\,\msun < M_L < 0.66\,\msun$, and we interpolate between the
\citet{delfosse00} and \citet{henry1999} relations for $0.10\,\msun < M_L < 0.12\,\msun$. 
The 99\%  and 95\% confidence level
upper limits on the lens system parameters from 2014 $I$ and $V$-band 
images are listed in Table \ref{tab-params-histogram-meas}. 
Since a detectable lens star must be close to the source in the galactic bulge, the dust 
in the foreground of the lens is similar to that of the source. Hence we use the same extinction 
for the lens and the source stars. For this upper limit, we assume a source distance 
$D_L = 8\,$kpc. This limit implies that the exoplanet is likely to be a sub-Saturn mass 
planet orbiting an M-dwarf star at a distance of $D_L \simlt 7.8\,$kpc toward 
the bulge.
    
\begin{deluxetable}{cccccccc}
\tablecaption{Triple Star Fit\label{tab-triple}}
\tablewidth{0pt}
\tablehead{\colhead{Year}&\colhead{Filter}&\multicolumn{3}{c}{Magnitude}&\colhead{Lens-Source}&\colhead{Angle}&\colhead{$\chi^2$} \\
& &Source & Lens & Blend Star&Separation (mas)&&}
\startdata
\multirow{2}{1 cm}{2012} &$I$ &19.62(0.05) & 22.62(0.44)&22.11(0.91) &17.6(1.5)&232.9(22.2)&206.1\\
& $V$&21.11(0.07) & 23.08(0.53)&22.71(0.83)&17.2(2.3) &84.2(49.7)&195.8\\
\multirow{2}{1 cm}{2014}&$I$ &19.58(0.04) &23.39(0.63) &21.93(0.77) &27.2(1.5)&74.7(38.9)&210.5\\
&$V$&20.94(0.08)&23.84(0.69)&22.35(0.78)&27.3(2.2)&68.8(57.6)&197.1\\
\enddata
\end{deluxetable}

\begin{deluxetable}{cccccc}
\tablecaption{Upper Limit Constraints on the Lens System Parameters\label{tab-params-histogram-meas}}
\tablewidth{0pt}
\tablehead{\colhead{parameter}&\colhead{units}&\multicolumn{2}{c}{99$\%$ confidence}&\multicolumn{2}{c}{95$\%$ confidence}\\
& & $I$&$V$&$I$&$V$}
\startdata
Host star mass, $M_*$&${\msun}$&0.64&0.73&0.61&0.72\\
Planet mass, $m_p$&$M_{\oplus}$&72&82&69&81\\
Host star - Planet 2D separation, $a_{\perp}$&AU&1.1&1.1&1.1&1.1\\
Lens distance, $D_L$&kpc &7.8&7.8&7.8&7.8\\
\enddata\\
$I_L \geq 22.15, V_L \geq 23.41$ (99$\%$ confidence)\\
$I_L \geq 22.44, V_L \geq 23.62$ (95$\%$ confidence)
\end{deluxetable}   

\begin{deluxetable}{cccc}
\tablecaption{The Blend Star Position from 3-Star PSF Fits
\label{tab-triple_pos}}
\tablewidth{0pt}
\tablehead{Year&Filter&\multicolumn{2}{c}{Source- Blend Star Separation}\\
&&x-direction (mas)&y-direction (mas)}
\startdata
2012&$I$&11.6(10.4)&12(10.4)\\
&$V$&8.8(7.2)&12.4(8.4)\\
2014&$I$&12.4(10.4)&5.6(4.8)\\
&$V$&11.6(8.8)&6.8(5.6)\\
\enddata
\end{deluxetable}

%

\section{Color Dependent Centroid Shift}
\label{sec-col_cen_shift}

The lens-source separation can also be detected by the color dependent centroid shift method,
which has been used for planetary events OGLE-2003-BLG-235 \citep{bennett06} and 
OGLE-2005-BLG-071 \citep{dong-ogle71}. This method is only effective if the source and lens 
have different colors so that the centroid of the blended lens+source flux will be different in the
two passbands, $I$ and $V$ in our case. From the source flux constrained fit discussed
in Section \ref{sec-dual-psf} and presented in Table \ref{tab-dual-fit}, we see that that the ratio of the
blend flux to that of the source is $\sim$0.85 and $\sim$0.78 in the $I$ and $V$-bands, 
respectively. First, let us consider the case of this source flux constrained model. It implies that
the blend star is $\sim 13.8\,$mas away from the source. This implies that the centroid of the 
source and the blend star is $2.11\,$mas and $3.02\,$mas away from the source star in the
$I$ and $V$-bands, respectively. Hence, the centroid shift is $0.91\,$mas in this case. 

Now, let us consider the blend = lens situation, which has already been ruled out by the
two-star fits. In this case the blend-source separation will be $27\,$mas. This implies
a centroid shift or $4.12\,$mas and $5.91\,$mas away from the lens position in the
$I$ and $V$-bands, respectively. So, in this case, the color dependent centroid shift 
is $1.79\,$mas, in this blend = lens situation. 
 
With these theoretical calculations in hand, we proceed to measure the color dependent 
centroid shift for an additional test of the blend = lens scenario. We selected 20 isolated stars 
within 200 pixel radius of the target. These stars are selected to be within 0.1 magnitude of the 
color and 0.5 magnitude of $V$-band brightness of the target star. We use these stars
to measure the average centroid shifts of single stars between $I$ and $V$ band frames. Then,
we measure the centroid shift of the target star relative to these reference stars. 
We also compare the centroid shift of the target star to the centroid shift of each of the reference 
stars in each individual image. The measured centroid shift of the target star is 
$0.45 \pm 0.49\,$mas. The uncertainty is calculated from the RMS of the centroid shifts of the
reference stars. This measurement is within 1-$\sigma$ of the $0.91\,$mas centroid shifted
predicted by the flux constrained 2-star model, but it is 2.7-$\sigma$ less than the separation
expected from the blend = lens model. So, the blend = lens model is also ruled out by
this color dependent centroid shift test.

\section{The Blend Star Is {\it NOT} The Lens; What Is It Then?}
\label{sec-NOTlens}

The source flux constrained dual star fit to the 2014 data 
are not consistent with the blend = lens model because the measured source--blend
separation is much smaller than the predicted separation between the source and 
lens stars, and our attempt to detect the color dependent centroid shift tends to
confirm this conclusion. Without a clear detection of the lens star, we cannot 
use mass-luminosity relations to get a precise distance and masses for the host star
and planet, as we have previously done for OGLE-2005-BLG-169 \citep{ogle169,batistaogle169}.
We can still estimate the masses and distance to the lens system with Bayesian analysis,
using the Galactic model of \citet{bennett14},
but without a lens brightness measurement, this estimate has a significant dependence on 
our prior assumptions. We are interested in this event because it indicates the 
presence of an exoplanet, but we do not know if the probability of hosting a planet
similar to the detected planet depends on the host star mass or distance. 
The simplest assumption is that the probability that a lens star hosts 
a planet like MOA-2008-BLG-310Lb does not depend on the host star mass or
distance. We certainly know that planets have been discovered orbiting a wide
variety of stars, but we don't have strong information to indicate that the planet
hosting probability doesn't depend on the host star mass or distance. So, we
aren't sure how accurate this assumption is.

This Bayesian analysis indicates that the lens
star has a median $I$-band magnitude of $I_L \sim$ 26.2 with a 1-$\sigma$ range of
24.1-28.6 and a 2-$\sigma$ range of 23.0-36.1. (Note that this 2-$\sigma$ upper limit
of 36.1 is not a real magnitude. It is a magnitude that we assigned to brown
dwarfs, which are too faint to detect. $I_L \sim$ 26.2 is certainly too faint to 
detect. $I \sim$ 23.0 is enough to perturb the fit, while $I \sim$ 24.1 is starting
to become a bit marginal. From out MCMC analysis (assuming that all stars
are equally likely to host a planet), we find that 78$\%$ of the time, the lens
star is $< 1\%$ of the source brightness (that is $I_L > 24.58$). This implies
that is quite likely that the host star is too faint to have much influence on our
models of the {\it HST images}. If so, the source flux constrained fit listed in
Table \ref{tab-dual-fit} should provide a good model of this system.

The fit results in Table \ref{tab-dual-fit} indicate that the position of this blend star has 
consistent positions with respect to the source position in the two pass bands in epoch. 
It is quite possible that the lens is too faint to contribute significantly to the 
$I$ or $V$ band flux of the target. This would be the case if the lens is a faint M-dwarf or even 
a brown dwarf or white dwarf. If the lens star is too faint to be detectable, then there is likely to
be a single blend star that dominates the excess flux on top of the source in the $I$
and $V$ bands. This blend star could be a companion to the source or the lens or an 
unrelated star. According to the discovery paper, the {\it a priori} probability of a blend star within 0.5 
magnitudes of the $H$ band magnitude of the blend star for each of these possibilities $\sim 5 \%$.
The two epochs of observations yield the weighted mean of the separations of the blend star 
with respect to source in $I$ and $V$ bands are $\Delta x=10.7\pm 1.0\,$mas, 
$\Delta y=12.1\pm1.3\,$mas for 2012 and $\Delta x=12.0\pm 0.8\,$mas, 
$\Delta y=6.5\pm1.0\,$mas for 2014 data. This implies a relative proper motion of
this blend star with respect to the source of
$\bm{\mu}_{\rm rel,b,H}= (0.66\pm 0.65) \hat{x} - (2.84 \pm 0.83)\hat{y}\,$mas/yr from 
Table~\ref{tab-dual-fit}. The magnitude of the relative proper motion for this star is 
$\mu_{\rm rel,b,H} = 2.92\pm 0.83\,$mas/yr, which is consistent with the proper motion dispersion of the 
of bulge stars \citep{galactic_vel}. Hence, this star can be an unrelated nearby bulge star. 
The implied separation of the source and blend stars is $\geq129\pm 9\,$AU, while the 
relative proper motion implies a blend--source relative velocity of $\geq 111\pm 31$km/sec, which is
much larger than the $\sim 5\,$km/sec escape velocity implied by the source--blend separation.
So, the blend star cannot be a companion to the source if the lens star is faint.

The discovery paper indicates that the VLT/NACO AO data were taken 3.2 years before the
2012 {\it HST} images, so we can use the $\bm{\mu}_{\rm rel,b,H}$ above to estimate the separation
at the time of the high resolution VLT/NACO AO images. We determine a source--blend separation of
$\Delta x=8.6\,$mas, $\Delta y=21.2\,$mas at the time of the VLT images. So the source--blend 
separation was $22.9\,$mas, which is much smaller than the VLT image FWHM of $130\,$mas.
This explains why this blend star was not resolved in the VLT/NACO AO observations in 2008.

Since the magnitude uncertainties for the blend star are large, its color can vary anywhere 
from spectral type A to K, considering 2-$\sigma$ uncertainties on the magnitudes. But this
star has faint apparent magnitudes. So if this star is an A-star, it must be located very far 
behind the galactic bulge. But this would mean that it would be many scale heights below the
Galactic plane, where the density of A-stars is very low. So, it is reasonable to assume that this 
star is probably a G or K star residing in the galactic bulge. 

The final possibility is that this blend star is a companion to the lens star
\citep{bennett07,gould-2016}. The blend--source heliocentric relative proper motion is
$\mu_{\rm rel,b} = 2.92\pm 0.83\,$mas/yr, and this 2.3-$\sigma$ smaller than the lens--source
relative proper motion of $\mu_{\rm rel} = 4.98\pm 0.31\,$mas/yr. So, a companion to the lens
is marginally excluded, if the lens star has negligible brightness.

A companion
to the source or lens might be possible if the lens star contributes significantly to the flux. In that
case, we would need to use the triple star fit instead of the source flux constrained 2-star
fit. As Table~\ref{tab-triple_pos} indicates, the relative positions of the blend and source stars
are not well constrained in these models.

We can also use the 3-star fits also we can check if the blend star is a 
companion or an ambient star in the case where the lens star flux is not negligible. 
But from Table 6 we find that the blend--source relative proper motion is 
$\mubold_{\rm rel,H,b} = (0.6\pm 4.8) \hat{x} - (3.0 \pm 4.2)\hat{y}\,$mas/yr. 
Without a constraint that the lens star flux is negligible, the error bars on the 
blend--source relative proper motion are too large to constrain any of the possibilities:
an ambient star or a companion to the source or lens.

\begin{figure}
\epsscale{1.0}
\plotone{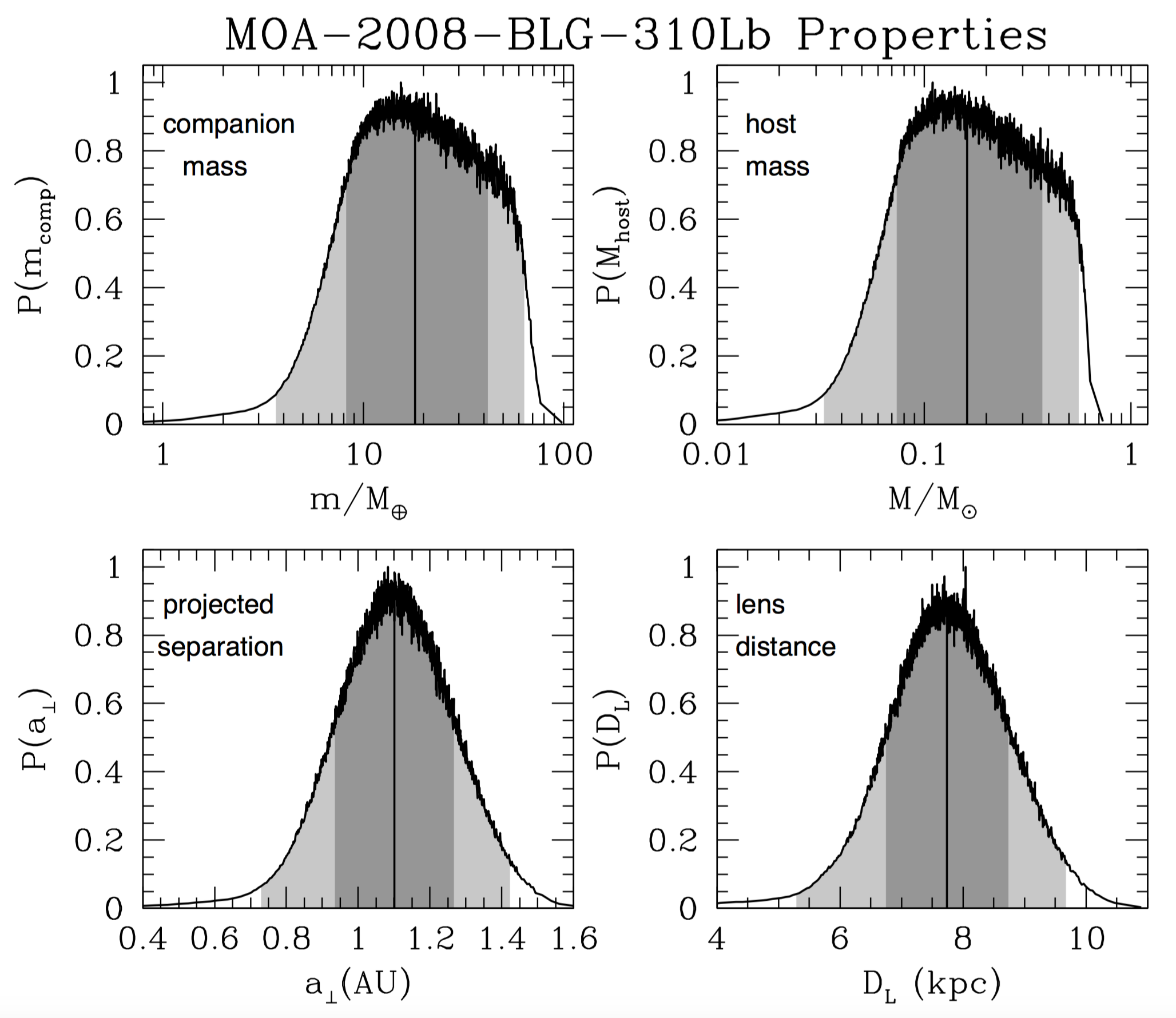}
\caption{The results of a Bayesian analysis showing the probability 
distributions for the planet and host star mass, planet - host star separation and lens distance. 
This analysis was done with a Markov Chain of light curve models constrained by our
{\it HST} upper limit on the host star brightness. It assumes that the probability of hosting
the detected planet does not depend on host star mass or distance.} 
\label{fig-histogram1}
\end{figure}

\begin{deluxetable}{cccc}
\tablecaption{Planetary System Parameters from Bayesian Analysis with Lens Flux Constraint\label{tab-params-histogram}}
\tablewidth{0pt}
\tablehead{\colhead{parameter}&\colhead{units}&\colhead{mean values \& RMS}&\colhead{2-$\sigma$ range}}
\startdata
Host star mass, $M_*$&${\msun}$&$0.21\pm 0.14$ & 0.033--0.56\\
Planet mass, $m_p$&$M_{\oplus}$& $23.4\pm 17$& 3.7--64\\
Host star - Planet 2D separation, $a_{\perp}$&AU&$1.10\pm 0.17$& 0.73--1.42\\
Host star - Planet 3D separation, $a_{3D}$&AU&$1.61\pm 0.98$& 0.82--4.75\\
Lens distance, $D_L$&kpc &$7.7\pm 1.1$& 5.3--9.7\\
Lens magnitude, $I_L$&Cousins $I$&$26.2\pm 2.2$& 23.0--36.1 \\
\enddata
\end{deluxetable}

\section{Discussion and Conclusion}
\label{sec-Conclusion}

As mentioned in Section~\ref{sec-NOTlens}, we have performed a Bayesian analysis to estimate 
the lens system properties using the Galatic model of \citet{bennett14} assuming that 
all the stars and brown dwarfs have an equal probability of hosting a planet with the
detected properties. We ran a series of Markov chains with a total of 662,000 links to
determine the allowed distribution of lens parameters, including the lens brightness 
constraint from Section~\ref{sec-triple-psf}, and the results of this calculation are
presented in Figure~\ref{fig-histogram1} and Table~\ref{tab-params-histogram}.

The source star brightness, color and the radius for each model in the Markov chain are 
included in this calculation. The source star is assumed to be a bulge star distributed at a 
distance $5\,{\rm kpc} \leq D_L \leq 12\,$kpc following a standard galactic model \citep{bennett14}. 
For each model, the source distance is chosen randomly from the microlensing rate
weighted galactic bulge distribution. We use empirical mass--luminosity relations 
\citep{henry1993, henry1999, delfosse00} and the mass--distance relations given in
equation~\ref{eq-m_thetaE} to determine the lens distance, $D_L$, the host star and the 
planet masses ($M_*$ and $m_p$), the host star $I$-band magnitude, $I_L$, and the 
host star--planet projected separation, $a_\perp$. The uncertainties shown in 
Table~\ref{tab-params-histogram} represent 1-$\sigma$ error bars. The mass of the host 
star (lens star) is approximately determined to be $M_* = 0.21^ {+0.21}_{-0.09} \msun$,
so it could be an M--dwarf of a brown dwarf. The predicted magnitude of the lens is
$I_L = 26.2^{+2.3}_{-2.1}$ with a 2-$\sigma$ range extending down to $I_L = 23.0$.
So, unless the lens brightness is near this 2-$\sigma$ upper limit, it will be too faint
to detect while blended with the $I_S = 19.43\pm 0.05$ source star. The planet mass is
$m_p = 23^{+24}_{-10} M_{\oplus}$, with a 2-$\sigma$ range of $3.7\mearth < m_p < 64 \mearth$,
so it could be a super-earth or a sub-saturn mass gas giant.
The distance to the lens system is more precisely determined, due to the relatively small
angular Einstein radius, $\theta_E = 0.29 \pm 0.05\,$mas. Our analysis predicts a
lens system distance of $D_L = 7.7 \pm 1.1\,$kpc. This implies that the lens system
is very likely to be in the galactic bulge, as claimed by the discovery paper \citep{moa310},
although a super-earth planet orbiting a brown dwarf in the disk cannot be ruled out. 

In this paper we have developed a technique for fitting a two or three star PSFs to the blended 
{\it HST} images consisting of a microlensed source, the lens star, and possible companions
to the source and lens stars, following a procedure outlined by 
\citet{bennett07}. We have constrained these fits with constraints on the source
flux and the lens--source separation from the microlensing light curve model. In our previous
analysis of event OGLE-2005-BLG-169 \citep{ogle169}, the unconstrained best fit solution 
was consistent with the source flux and the predicted lens--source separation in three
different passbands, as well as subsequent Keck AO $H$-band images \citep{batistaogle169}. 
Since the lens--source relative proper motion is determined by the planetary signature in the 
light curve, the confirmation of the separation predicted by the light curve is also a confirmation
of the planetary interpretation of the light curve. This confirmation was important for OGLE-2005-BLG-169 event due to the fragmentary character of the light curve over peak. But for MOA-2008-BLG-310 event, the light curve is well covered, allowing very little scope for incorrect modeling. Hence we can expect to measure the relative lens-source proper motion more precisely and compare with the lens-source proper motion derived from the light curve modeling. But the event
MOA-2008-BLG-310 did not yield a confirmation of the blend = lens model. Instead, we found
that the source flux constrained solutions were not consistent with the lens--source
relative proper motion predicted by the light curve. The two epochs of {\it HST} data were taken in 2012,
3.62 years after the event, and 2014, 5.59 years after the peak magnification. The expected 
lens--source separations at these two observing epochs are $17.4 \pm 0.4\,$mas and 
$27.4 \pm 0.7\,$mas, for the 2012 and 2014 observations, respectively. The 2014 {\it HST} observations,
in particular, were not consistent with the blend = lens model. The lens--source 
separation constrained models had a $\chi^2$ increase of $\Delta\chi^2 = 40.8$ compared to
models, which only constrained the source flux. So, we have concluded that the blend flux, first
identified by \citet{moa310} is {\it not} due to the planetary host (and lens) star. This conclusion
is strengthened by the lack of a detectable color dependent centroid shift \citep{bennett06}
Therefore, we consider 3-star models in order to constrain the brightness and mass of the planetary
host star. In these fits, we constrain the source flux and length, but not the direction of  the
lens--source separation vector.

As discussed in Sections~\ref{sec-dual-psf} and \ref{sec-triple-psf}, there are two possible 
solutions available. One is that the lens is too faint to influence the {\it HST} image models, 
and the extra flux is solely due to a star that is an unrelated nearby star. The second possibility 
is that the lens is bright enough ($I\sim 23$) to influence the ePSF fit. In this case, with two
additional stars, besides the source star, contributing flux, the uncertainty on the proper motion
of the blend star is much larger. So, the
blend star in the 3-star fit can be either a nearby unrelated star or a companion to the source or the lens.  
The triple star fit does provide an upper limit of the host star mass and therefore
the planet mass. This implies that the exoplanetary system is a sub Saturn mass system located 
in the bulge orbiting around a M--dwarf star or a brown dwarf. 

The method that we have presented in this paper is an important development in the effort to
characterize exoplanets found by the microlensing method. We have shown that it is possible
to determine whether excess flux at the location of a microlensed source star is due to the lens
star of a planetary microlensing event. In the case of OGLE-2016-BLG-169, such measurements
indicated that the blend star was the planetary host star \citep{ogle169}, but for MOA-2008-BLG-310, 
we have shown that the blend star is not the lens. Such measurements are possible for most
planetary microlensing events because most planetary microlensing features resolve the 
finite angular size of the source star and predict the lens--source relative proper 
motion, $\mu_{\rm rel}$ which allows the test that we have performed in this paper. 
A few years after the microlensing event when lens has moved away from the source, 
the lens and source stars will be partially resolved. Then, the analysis of the high resolution 
images of the partially resolved lens--source system will allow us to detect the lens (and planetary
host star) and determine its mass and that of its planet. The method presented in this paper allows
us to determine if any excess flux is actually due to the host star. 

This characterization of planetary systems discovered by microlensing is important because
microlensing is unique in its sensitivity to the cold low--mass exoplanets beyond the snowline, 
where other exoplanet detection methods are not so effective. But the light curve analysis of 
most planetary microlensing events yields only the planet--host star mass ratio and separation
in Einstein radius units. More observations like the ones we have analyzed in this
paper of MOA-2008-BLG-310 and those of OGLE-2005-BLG-169 \citep{ogle169} will
allow us to expand the current state of the art analysis of exoplanet statistics beyond
the snow line \citep{suzuki2016} to include the dependence of the exoplanet mass function
on the host mass and distance.

The analysis we have presented here confirms the prediction 
\citep{bennett02,bennett07} that such measurements should be possible for virtually all planets
found by a space-based microlensing survey. In fact, this method is likely to to be the 
primary method for determining planet and host star masses for exoplanets discovered
by the WFIRST exoplanet microlensing program \citep{WFIRST_AFTA}. The advantage
of a space-based microlensing survey is that the WFIRST observations themselves will
provide the high angular resolution observations needed to detect the exoplanet host stars.
However, the WFIRST fields will be more crowded than the fields of OGLE-2005-BLG-169
or MOA-2008-BLG-310, in part because WFIRST will observe in the infrared. Hence the 
probability of blending by unrelated stars will be higher. Thus, it will be more important to 
distinguish lens from unrelated blend stars and companions to the source and lens. Therefore, it
will be necessary to use the method developed in this paper to avoid errors in determining 
the host star brightness, mass and distance. About $44\%$ of G--dwarfs have stellar
companions as do 26\% of K and M--dwarfs \citep{duchene2013}. So stellar companions
will be a common source of contamination in the attempt to identify host stars for planets
discovered by the WFIRST microlensing program. With WFIRST it will be more the rule
than the exception to have source, lens and an additional stellar companion that is faint.
Hence getting the excess flux is probably not enough in WFIRST era, we will have to follow a more careful investigation, like MOA-2008-BLG-310 follow up analysis. Thus, the methods that we have developed
in this paper are an important step forward in the development of the WFIRST exoplanet mass 
measurement method that will measure the demographics of cool and cold exoplanets. 

Based on observations made with the NASA/ESA Hubble Space Telescope, 
obtained at the Space Telescope Science Institute (STScI), which is operated by the 
Association of Universities for Research in Astronomy, Inc., under NASA contract 
NAS 5-26555. These observations are associated with programs \# 12541 and 13417.A.B. and D.P.B. 
were supported by NASA through grants NASA-NNX12AF54G and 
NNX13AF64G. I.A.B. was supported by the Marsden Fund of Royal Society of 
New Zealand, contract no. MAU1104.

\end{document}